\documentclass[aps,prb,twocolumn,showpacs,groupedaddress]{revtex4-2}
\usepackage{graphicx}
\usepackage{epstopdf} 
\usepackage{bm}
\usepackage{amsmath}
\begin{document}

\title{\emph{Ab initio} analytical model of lasing in plasmonic lattices}

\author{V.G.~Bordo}
\email{vgbordo@gmail.com}
\affiliation{Independent Researcher, DK-6400 S{\o}nderborg, Denmark}


\date{\today}

\begin{abstract}
The first-principles theory of lasing in a rectangular lattice of spherical metal nanoparticles is developed in a fully analytical form in the dipole approximation. The lasing conditions are obtained for different diffraction orders, both propagating and evanescent. Their analysis reveals that besides usual lasing there can be lasing without population inversion which is invisible in conventional experiments, but can be observed in total internal reflection.
\end{abstract}


\maketitle

\section{Introduction} 
Recently there has been growing interest to regular arrays of metal nanoparticles (NPs) which received the name of plasmonic lattices. They can support the so-called plasmonic surface lattice resonances (SLRs) which are diffractively coupled plasmon resonances of individual NPs \cite{Kravets18}. Due to their extremely narrow spectral widths they are very promising for diverse applications, including ultrasensitive biodetection, improved photovoltaic cells, new optoelectronic and communication devices, more efficient photocatalysis and many others.\\ 
A particularly attractive applied direction is the development of lasers based on SLRs in view of low lasing thresholds stemming from very high quality factors of SLRs which resolves a formidable problem in plasmonics \cite{Wang18}. Lasing in plasmonic lattices were demonstrated in a variety of structures besides square lattices, for example at the high-symmetry points of the Brillouin zone in honeycomb and hexagonal structures \cite{Torma19,Juares22,Mattei23}, symmetry-breaking nanocrescent arrays \cite{Lin19}, supercell arrays \cite{Heilmann23}, incommensurate moir\'e lattices \cite{Fasanelli24}. The laser parameters allow great tunability and can be controlled through the dielectric environment \cite{Yang15}, NPs size \cite{Li19}, linear dimensions of the NP array \cite{Wang20}, incorporation of photochromic molecules into the liquid gain medium \cite{Taskinen20}, mixed dye solution \cite{Guan23}.\\
The complexity of the phenomena which govern the operation of plasmonic lattice lasers requires an adequate theoretical description. A conventional approach is based on large-scale numerical calculations which are capable to account the details of the light-matter interaction. In its advanced formulation, the approach involves numerical solution of coupled Maxwell-Liouville equations which describe the interaction between the gain medium and the plasmonic lattice \cite{Trivedi17}. Although such a method can account for subtle features of the system evolution, its results cannot be transferred to another, even similar, system with a different set of parameters. On the contrary, a rigorous analytical model, if available, does not focus on insignificant details, while being able to follow the general dependencies.\\
In the present paper, we develop a rigorous theory of lasing in the molecular medium near a plasmonic lattice using the exact solution for the field emitted by a point oscillating dipole. We derive the lasing conditions for both $s$ and $p$ polarized emission in a simple analytical form. As we conclude from their analysis, besides the usual lasing which occurs under optical pumping, there can be lasing without population inversion. The latter effect is invisible in conventional experiments and has not been discussed in the literature on lasing in plasmonic lattices before.\\
The paper is organized as follows. In Sec. \ref{sec:model} the theoretical model adopted in the paper is introduced. In Secs. \ref{sec:local} and \ref{sec:evolution} the equations for the local field in the gain medium and for the gain medium polarization, respectively, are derived. Section \ref{sec:lasing} deals with the criterion for lasing and its analysis. Conclusion summarizes the main results of the paper.
\section{Theoretical model}\label{sec:model}
We consider a two-dimensional rectangular plasmonic lattice with the lattice constants $a$ and $b$ along the coordinate axes $x$ and $y$, respectively, composed of spherical metal NPs of radius $R$ with the dielectric susceptibility $\epsilon_m$. We assume that the space both above the lattice ($z>0$) and below it ($z<0$) is occupied by the host dielectric material which has the dielectric susceptibility $\epsilon_h$.\\ 
We assume that the dye molecules which constitute the gain medium fill the slab between the planes $z=0$, where the plasmonic lattice is located, and $z=h$ with the volume number density $N$. We model the molecules by two-level quantum systems with the transition frequency $\omega_0$ and the transition dipole moment ${\bm \mu}$.\\
In the quasistatic approximation, which is valid for the operating wavelengths $\lambda$ such that $\lambda\gg R$, the polarizability of a single isolated NP is found as (Gaussian units) \cite{Stratton}
\begin{equation}\label{eq:static}
\alpha_0(\omega)=\epsilon_hR^3\frac{\epsilon_m(\omega)-\epsilon_h}{\epsilon_m(\omega)+2\epsilon_h},
\end{equation}
where the frequency dependence of $\epsilon_m$ is given by
\begin{equation}
\epsilon_m(\omega)=\epsilon_{\infty}-\frac{\omega_P^2}{\omega(\omega-i\Gamma)}
\end{equation}
with $\epsilon_{\infty}$ the offset which takes into account the interband transitions, $\omega_P$ the plasma frequency of a metal and $\Gamma$ the relaxation constant. For larger NPs, the approximation (\ref{eq:static}) can be modified to account for dynamic polarization and radiative damping \cite{Barnes08}.\\
In view of the inequalities $\Gamma, \omega\ll \omega_P$, Eq. (\ref{eq:static}) can be approximated as
\begin{equation}
\alpha_0(\omega) \approx \epsilon_hR^3\frac{\omega_{SP}^2}{\omega_{SP}^2-\omega^2+i\Gamma\omega},
\end{equation}
where
\begin{equation}
\omega_{SP}=\frac{\omega_P}{\sqrt{\epsilon_{\infty}+2\epsilon_h}}
\end{equation}
is the frequency of the NP surface plasmon. This expression is further simplified in the vicinity of the resonance where $\mid\omega-\omega_{SP}\mid\ll \omega_{SP}$ as follows
\begin{equation}\label{eq:alpha0r}
\alpha_0(\omega) \approx -\frac{A}{\omega-\omega_{SP}-i(\Gamma/2)}
\end{equation}
with $A=\epsilon_hR^3\omega_{SP}/2$.\\
In an infinite lattice, the quantity (\ref{eq:static}) should be replaced by an effective polarizability \cite{Barnes08}
\begin{equation}
\alpha(\omega)=\frac{1}{1/\alpha_0(\omega) - S},
\end{equation}
where the dipole sum $S=S^{\prime}+iS^{\prime\prime}$ accounts for the influence of all other lattice NPs in the dipole approximation. Substituting the expression (\ref{eq:alpha0r}) into this equation one obtains
\begin{equation}
\alpha(\omega)\approx -\frac{A}{\omega - \omega_{SLR}-i\Gamma_{SLR}},
\end{equation}
where
\begin{equation}
\omega_{SLR}=\omega_{SP}-AS^{\prime}
\end{equation}
and
\begin{equation}
\Gamma_{SLR}=\frac{\Gamma}{2}-AS^{\prime\prime}
\end{equation}
are the frequency of the plasmonic SLR and its width, respectively \cite{Kravets18}.\\
\section{Local field in the gain medium}\label{sec:local}
Let us assume that there is an external electric field ${\bf E}_0({\bf r},t)={\bf E}_0({\bf r})\exp(-i\omega t)$ and consider the local field in the gain medium which governs the evolution of the dye molecules polarization, ${\bf P}$. Besides the well-known Lorentz field
\begin{equation}
{\bf E}_L({\bf r})={\bf E}_0({\bf r})+\frac{4\pi}{3\epsilon_h}{\bf P}({\bf r}),
\end{equation}
it contains a contribution originating from the plasmonic lattice backaction \cite{Bordo22}, ${\bf E}_{pl}$, which can be found as follows.\\
The electric field generated at point ${\bf r}=({\bm\rho},z)$ by a point dipole ${\bm \mu}$ oscillating with the frequency $\omega$ and representing a dye molecule located at point ${\bf r}_0=({\bm\rho}_0,z_0)$ can be conveniently written in the form
\begin{equation}\label{eq:dipole}
{\bf E}({\bf r})=\bar{\bf F}({\bf r}-{\bf r}_0)\cdot{\bm \mu},
\end{equation}
where $\bar{\bf F}$ is the so-called field susceptibility tensor. The latter quantity has the form of a two-dimensional Fourier integral \cite{Sipe81}
\begin{equation}
\bar{\bf F}({\bf r}-{\bf r}_0)=\int \frac{d{\bm\kappa}}{(2\pi)^2}\bar{\bf f}({\bm \kappa};z-z_0)\exp[i{\bm \kappa}\cdot ({\bm \rho}-{\bm\rho}_0)], 
\end{equation}
where
\begin{eqnarray}
\bar{\bf f}({\bm \kappa};z-z_0)=2\pi i\frac{\tilde{\omega}^2}{W_h}\left[(\hat{s}\hat{s}+\hat{p}_+\hat{p}_+)\theta(z-z_0)e^{iW_h(z-z_0)}\right.\nonumber\\
+\left.(\hat{s}\hat{s}+\hat{p}_-\hat{p}_-)\theta(z_0-z)e^{-iW_h(z-z_0)}\right]-4\pi\hat{z}\hat{z}\delta(z-z_0).\nonumber\\
\end{eqnarray}
Here $\tilde{\omega}=\omega/c=2\pi/\lambda$ with $c$ being the speed of light in vacuum, $W_h=(\tilde{\omega}^2\epsilon_h-\kappa^2)^{1/2}$, the unit vectors $\hat{z}$, $\hat{\kappa}$, and $\hat{s}=\hat{\kappa}\times\hat{z}$ are oriented along the corresponding directions, $\theta(Z)$ is the unit step function, and
\begin{equation}
\hat{p}_{\pm}=\frac{1}{\tilde{\omega}\sqrt{\epsilon_h}}(\kappa\hat{z}\mp W_h\hat{\kappa}).
\end{equation}
The field (\ref{eq:dipole}) induces in its turn the dipole moment of the metal NP located at point ${\bf r}_i=({\bm\rho}_i,0)$ given by
\begin{equation}
{\bf p}_i=\alpha(\omega) {\bf E}({\bf r}_i)=\alpha(\omega)\bar{\bf F}({\bf r}_i-{\bf r}_0)\cdot{\bm\mu}.
\end{equation}
Then the field generated by this NP at the observation point ${\bf r}$ is found as
\begin{equation}
{\bf E}_i({\bf r})=\bar{\bf F}({\bf r}-{\bf r}_i)\cdot{\bf p}_i=\alpha(\omega)\bar{\bf F}({\bf r}-{\bf r}_i)\bar{\bf F}({\bf r}_i-{\bf r}_0)\cdot{\bm\mu}.
\end{equation}
The overall contribution from all dye molecules and all NPs is obtained by integration over the gain medium volume $V$ and summation over the lattice as
\begin{equation}\label{eq:field}
{\bf E}_{pl}({\bf r})=\alpha(\omega)\sum_i\bar{\bf F}({\bf r}-{\bf r}_i)\int_V \bar{\bf F}({\bf r}_i-{\bf r}^{\prime})\cdot{\bf P}({\bf r}^{\prime})d{\bf r}^{\prime},
\end{equation}
where we have used that ${\bf P}=N{\bm \mu}$.\\
Due to the periodicity of the plasmonic lattice both the field (\ref{eq:field}) and the polarization are periodic as well and can be expanded in the Fourier series as follows
\begin{equation}
{\bf E}_{pl}({\bf r}) =\sum_{mn}{\bf e}_{mn}(z)\exp\left(i\frac{2\pi m}{a}x\right)\exp\left(i\frac{2\pi n}{b}y\right),
\end{equation}
\begin{equation}
{\bf P}({\bf r}) =\sum_{mn}{\bf p}_{mn}(z)\exp\left(i\frac{2\pi m}{a}x\right)\exp\left(i\frac{2\pi n}{b}y\right),
\end{equation}
where the Fourier coefficients ${\bf e}_{mn}$ are found as 
\begin{eqnarray}
{\bf e}_{mn}(z)=\frac{1}{ab}\int_{-b/2}^{b/2}\int_{-a/2}^{a/2}{\bf E}({\bf r})\exp\left(-i\frac{2\pi m}{a}x\right)\nonumber\\
\times\exp\left(-i\frac{2\pi n}{b}y\right)dxdy
\end{eqnarray}
and the Fourier coefficients ${\bf p}_{mn}$ are found analogously.\\
Substituting here the expression (\ref{eq:field}) and using the representation
\begin{equation}
\delta(X)=\frac{1}{2\pi}\sum_{n=-\infty}^{\infty}e^{inX},
\end{equation}
one obtains the relation between the Fourier coefficients ${\bf e}_{mn}$ and ${\bf p}_{mn}$ as follows
\begin{eqnarray}
{\bf e}_{mn}(z)=\frac{\alpha(\omega)}{ab}\bar{\bf f}\left(\frac{2\pi m}{a},\frac{2\pi n}{b};z\right)\nonumber\\
\times\int_0^h\bar{\bf f}\left(\frac{2\pi m}{a},\frac{2\pi n}{b};-z^{\prime}\right){\bf p}_{mn}(z^{\prime})dz^{\prime}.
\end{eqnarray}
\section{Evolution of the gain medium polarization}\label{sec:evolution}
We represent the time dependence of both the local field and the polarization in the forms
\begin{equation}
{\bf E}_{loc}({\bf r},t)=\tilde{\bf E}({\bf r},t)\exp(-i\omega t)
\end{equation}
and
\begin{equation}
{\bf P}({\bf r},t)=\tilde{\bf P}({\bf r},t)\exp(-i\omega t),
\end{equation}
respectively, where tilde denotes the slowly varying in time amplitudes.\\
For the field amplitudes far below the saturation field and in the rotating wave approximation the polarization of the dye molecules obeys the optical Bloch equation \cite{Haken}
\begin{equation}\label{eq:Bloch}
\frac{\partial\tilde{\bf P}({\bf r},t)}{\partial t}=-(\gamma_{\perp}-i\Delta)\tilde{\bf P}({\bf r},t)-\frac{i}{3\hbar}\mu^2D_0\tilde{\bf E}({\bf r},t),
\end{equation}
where $\gamma_{\perp}$ is the transverse (phase) relaxation rate, $\Delta=\omega-\omega_0$ and $D_0=Nw_0$ with $w_0$ being the equilibrium value of the population inversion between the upper and lower quantum states of a molecule which is determined by optical pumping.\\
Expanding Eq. (\ref{eq:Bloch}) in the Fourier series, one obtains
\begin{eqnarray}\label{eq:Bloch1}
\frac{\partial{\bf p}_{mn}(z,t)}{\partial t}=-(\gamma_{\perp}-i\Delta){\bf p}_{mn}(z,t)\nonumber\\
-\frac{i}{3\hbar}\mu^2D_0\left[{\bf e}_{mn}^0(z)+\frac{4\pi}{3\epsilon_h}{\bf p}_{mn}(z,t)+{\bf e}_{mn}(z,t)\right],
\end{eqnarray}
where ${\bf e}_{mn}^0$ are the Fourier coefficients of the external field. Now, introducing the Laplace transforms in time 
\begin{equation}
{\bf e}_{mn}(s)=\int_0^{\infty}{\bf e}_{mn}(t)e^{-st}dt
\end{equation}
and
\begin{equation}
{\bf p}_{mn}(s)=\int_0^{\infty}{\bf p}_{mn}(t)e^{-st}dt
\end{equation}
with the initial conditions ${\bf p}_{mn}(t=0)=0$, one comes to the set of integral equations
\begin{eqnarray}\label{eq:p}
s{\bf p}_{mn}(z,s)=-(\gamma_{\perp}-i\Delta^{\prime}){\bf p}_{mn}(z,s)\nonumber\\
-\frac{i}{3\hbar}\mu^2D_0\left[\frac{{\bf e}_{mn}^0(z)}{s}+{\bf e}_{mn}(z,s)\right]
\end{eqnarray}
\begin{eqnarray}\label{eq:e}
{\bf e}_{mn}(z,s)=\frac{\alpha(\omega)}{ab}\bar{\bf f}\left(\frac{2\pi m}{a},\frac{2\pi n}{b};z\right)\nonumber\\
\times\int_0^h \bar{\bf f}\left(\frac{2\pi m}{a},\frac{2\pi n}{b};-z^{\prime}\right){\bf p}_{mn}(z^{\prime},s)dz^{\prime},
\end{eqnarray}
where 
\begin{equation}
\Delta^{\prime}=\Delta - \frac{4\pi\mu^2D_0}{9\epsilon_h\hbar}.
\end{equation}
Let us introduce a new quantity
\begin{equation}
\vec{\cal P}_{mn}(s)=\int_0^h \bar{\bf f}\left(\frac{2\pi m}{a},\frac{2\pi n}{b};-z\right){\bf p}_{mn}(z,s)dz.
\end{equation}
Then from Eqs. (\ref{eq:p}) and (\ref{eq:e}) one obtains a vector equation 
\begin{equation}\label{eq:matrix}
\left[\bar{M}_{mn}-\lambda (s)\bar{I}\right]\vec{\cal P}_{mn}(s)=-\frac{1}{s}\vec{\cal E}_{mn},
\end{equation}
where 
\begin{eqnarray}
\bar{M}_{mn}=\frac{\alpha(\omega)}{ab}\nonumber\\
\times\int_0^h \bar{\bf f}\left(\frac{2\pi m}{a},\frac{2\pi n}{b};-z\right)\bar{\bf f}\left(\frac{2\pi m}{a},\frac{2\pi n}{b};z\right)dz,
\end{eqnarray}
\begin{equation}
\vec{\cal E}_{mn}=\int_0^h \bar{\bf f}\left(\frac{2\pi m}{a},\frac{2\pi n}{b};-z\right){\bf e}_{mn}^0(z)dz,
\end{equation}
\begin{equation}
\lambda^{-1}(s)=-\frac{i}{3\hbar}\frac{\mu^2D_0}{s+\gamma_{\perp}-i\Delta^{\prime}}
\end{equation}
and $\bar{I}$ is a unit tensor of rank $3$. Equation (\ref{eq:matrix}) governs the evolution of the molecular polarization as well as the local field in the gain medium.
\section{Lasing condition}\label{sec:lasing}
The time dependence of both the polarization and the field in the diffraction orders specified by the integers $m$ and $n$ is given by the poles $s_{mn}$ of the Laplace transformed quantities ${\bf p}_{mn}(z,s)$. The poles in their turn are determined by the eigenvalues of the tensors $\bar{M}_{mn}$, $\lambda_k^{mn}$ ($k=1,2,3$), through the equation $\lambda(s_{mn})=\lambda_k^{mn}$. The lasing threshold is found from the condition that at least one of the poles lies on the imaginary axis of the complex plane of $s$ and the further change of the system parameters moves it to the right half-plane of the complex plane. Representing the poles as $s_{mn}=\sigma_{mn}+i\Omega_{mn}$, one can write this condition as $\sigma_{mn}>0$ that corresponds to the exponential increase in time proportional to $\exp(\sigma_{mn}t)$ and signifies lasing.\\
For the further analysis, it is convenient to split the problem into two parts, one of which considers the field components parallel to the plane of the plasmonic lattice ($s$ polarization) and the other one deals with the out-of-plane components ($p$ polarization).
\subsection{$s$ polarization}
For $s$ polarization the matrix of the tensor $\bar{M}_{mn}$ is reduced to a constant
\begin{equation}\label{eq:Mmn}
M_{mn}^s=-i\frac{2\pi^2\alpha(\omega)\tilde{\omega}^4}{abW_{mn}^3}[1-\exp(2iW_{mn}h)],
\end{equation}
where 
\begin{equation}\label{eq:Wmn}
W_{mn}=(\tilde{\omega}^2\epsilon_h - \kappa_{mn}^2)^{1/2}
\end{equation}
with
\begin{equation}\label{eq:kappa}
\kappa_{mn}=\left[\left(\frac{2\pi m}{a}\right)^2+\left(\frac{2\pi n}{b}\right)^2\right]^{1/2}.
\end{equation}
Correspondingly, $\lambda(s_{mn})=M_{mn}^s$ and one finds from here
\begin{equation}
s_{mn}=-\gamma_{\perp}+i\Delta^{\prime}-i\frac{\mu^2D_0}{3\hbar}M_{mn}^s
\end{equation}
and the lasing condition
\begin{equation}\label{eq:lasing_s}
\frac{\mu^2D_0}{3\hbar}\text{Im}(M_{mn}^s)>\gamma_{\perp}.
\end{equation}
It has a clear physical sense: the excitation rate given by the left-hand side should exceed the phase relaxation rate.\\
The above derived condition concerns, however, a two-level model of dye molecules. Real molecules possess broad absorption and emission bands which originate from multiple electronic-vibrational transitions of a molecule. Then the field generation can be regarded as multimode lasing which is fed from a common source. It is well known that in such a situation only the mode which has the largest excitation rate grows, while the other modes are discriminated \cite{Haken}. We assume therefore that lasing occurs at $\omega\approx\omega_{SLR}$ where the polarizability $\alpha(\omega)$ is maximum in magnitude and equals $\alpha(\omega_{SLR})=-i\epsilon_hR^3Q$ with $Q=\omega_{SLR}/(2\Gamma_{SLR})$ being the quality factor of the SLR.\\
In what follows, for the sake of simplicity, we focus our attention on square plasmonic lattices for which $b=a$. Then the quantity $W_{mn}$, which represents the wave vector $z$-component of the field generated in the diffraction order $(m,n)$, takes the form $W_{mn}=\epsilon_h^{1/2}\tilde{\omega}w_{mn}$ with
\begin{equation}
w_{mn}=\left[1-\frac{1}{\epsilon_h}\left(\frac{\lambda}{a}\right)^2(m^2+n^2)\right]^{1/2}.
\end{equation}
It is reasonable to consider the following two cases.\\
(i) $w_{mn}$ is real. This condition limits the range of the diffraction orders which can be observed in lasing as propagating (radiative) modes as follows
\begin{equation}
m^2+n^2 <\epsilon_h \left(\frac{a}{\lambda}\right)^2.
\end{equation}
In other words, on the plane of integers $(m,n)$ such diffraction orders are located inside the circle of radius $\epsilon_h^{1/2}(a/\lambda)$. In this case
\begin{equation}
\text{Im}(M_{mn}^s)=4\pi^2\frac{R^3hQ\tilde{\omega}^2}{a^2w_{mn}^2}\frac{\sin\xi_{mn}}{\xi_{mn}}
\end{equation}
with $\xi_{mn}=2\epsilon_h^{1/2}\tilde{\omega}w_{mn}h$. \\
(ii) $w_{mn}$ is imaginary. The corresponding diffraction orders correspond to evanescent (non-radiative) modes which propagate along the palsmonic lattice and satisfy the inequality
\begin{equation}
m^2+n^2 >\epsilon_h \left(\frac{a}{\lambda}\right)^2.
\end{equation}
Making the substitution $w_{mn}=iu_{mn}$ with $u_{mn}$ being a real and positive quantity, one finds
\begin{equation}\label{eq:evan}
\text{Im}(M_{mn}^s)=-2\pi^2\frac{R^3Q\tilde{\omega}}{\epsilon_h^{1/2}a^2u_{mn}^3}\left[1-\exp(-2\epsilon_h^{1/2}\tilde{\omega}u_{mn}h)\right].
\end{equation}
The quantity (\ref{eq:evan}) is definitely negative and the lasing condition (\ref{eq:lasing_s}) can only be satisfied if $D_0$ is negative as well, i.e. if the population inversion does not occur. This can be realized when there is a weak incoherent external electromagnetic field tuned to the absorption band of dye molecules. The resulting lasing can be observed in total internal reflection, for example when the structure is placed onto a prism with the refractive index $n_p>\epsilon_h^{1/2}$. Such lasing without inversion was also predicted before for lasing near a flat metal surface \cite{Bordo22} and near a random array of metal NPs \cite{Bordo21}.\\
The above results can be illustrated by experimental observations. For example, for a lattice of Au NPs with $a=600$ nm and $\epsilon_h^{1/2}=1.5$ there was observed lasing at $\lambda\approx 900$ nm along the normal to the lattice \cite{Yang15}. In such a case $\epsilon_h(a/\lambda)^2\approx 0.67$ and the radiative modes can only lase if $m=n=0$, i.e. at the $\Gamma$-point where the wave vector component parallel to the lattice is zero. All other lasing diffraction orders are evanescent and can only be observed in the absorption band in total internal reflection.
\subsection{$p$ polarization}
For $p$ polarization the matrix of the tensor $\bar{M}_{mn}$ can be written in the form
\begin{equation}
\hat{M}_{mn}^p=M_{mn}^s\frac{W_{mn}^2-\kappa_{mn}^2}{\epsilon_h^2\tilde{\omega}^4}
\begin{pmatrix}\label{eq:Mp}
W_{mn}^2 & \kappa_{mn}W_{mn}\\
-\kappa_{mn}W_{mn} & -\kappa_{mn}^2
\end{pmatrix},
\end{equation}
where the quantities $M_{mn}^s$, $W_{mn}$ and $\kappa_{mn}$ are defined by Eqs. (\ref{eq:Mmn}), (\ref{eq:Wmn}) and (\ref{eq:kappa}), respectively. One of the eigenvalues of the matrix (\ref{eq:Mp}) is zero which does not change anyhow the decay of the dye molecules polarization. The other eigenvalue equals
\begin{equation}\label{eq:Mmnp}
M_{mn}^p=\eta_{mn}M_{mn}^s
\end{equation}
with
\begin{equation}
\eta_{mn}=\frac{(W_{mn}^2-\kappa_{mn}^2)^2}{\epsilon_h^2\tilde{\omega}^4}
\end{equation}
and the lasing condition has the form
\begin{equation}\label{eq:lasing_p}
\frac{\mu^2D_0}{3\hbar}\eta_{mn}\text{Im}(M_{mn}^s)>\gamma_{\perp}.
\end{equation}
Due to the definite positiveness of the factor $\eta_{mn}$, the analysis of the lasing condition for radiative vs non-radiative diffraction orders is the same as for $s$ polarization. For lasing at the $\Gamma$-point $\kappa_{00}=0$, $\eta_{00}=1$ and the lasing thresholds for both $s$ and $p$ polarizations coincide with each other as expected. For the other radiative modes $\eta_{mn}<1$, whereas for the evanescent modes $\eta_{mn}>1$. This implies that there is an interval between the two lasing thresholds where the generated field has a certain polarization even when being fed by non-polarized light.
\section{Conclusion}
In the present paper, we have developed a first-principal approach to the dynamics of dye molecules in the vicinity of a plasmonic lattice. We modeled the NPs in the lattice by point dipoles which possess the polarizability of a metallic sphere, while the dye molecules were modeled by two-level quantum systems which obey the optical Bloch equations. The evolution of the system was found in terms of the Laplace transform of the molecular polarization which satisfies an integral equation. The expansion of this quantity in the two-dimensional Fourier series determined the evolution of individual diffraction orders. The lasing condition was identified as a shift of the Laplace transform pole to the right half-plane of the complex plane.\\
The criterion of lasing was obtained in a simple analytical form and analyzed for both $s$ and $p$ polarizations of the generated field. It was shown that lasing can occur for radiative diffraction orders as conventional lasing and for non-radiative orders as lasing without population inversion. The latter one manifests itself in generation of evanescent waves which can only be observed in total internal refection.\\
The experimental verification of lasing without optical pumping would open new possibilities for ultrasensitive surface analysis and development of nanoscale lasers which are fed by a weak incoherent electromagnetic field.

\end{document}